\documentclass[preprint2,twocolumn,times,tighten]{aastex63}

\usepackage{graphicx,times}
\usepackage{subfigure}
\newcommand{\be}{\begin{equation}}
\usepackage{threeparttable}
\usepackage{booktabs}
\newcommand{\ee}{\end{equation}}
\newcommand{\bea}{\begin{eqnarray}}
\newcommand{\eea}{\end{eqnarray}}

\usepackage{amsmath}
\usepackage{cases}
\usepackage{longtable}
\usepackage{hyperref}
\usepackage{epstopdf}
\usepackage{amsmath,bm}
\usepackage{amssymb}
\usepackage{natbib}
\usepackage{morefloats}
\usepackage{multirow}
\usepackage{array}
\usepackage{verbatim}
\usepackage{color}
\usepackage{lineno}
\begin{document}

\title{Quasi-perpendicular shock acceleration and TDE radio flares}

\author[0000-0002-0458-7828]{Siyao Xu}
\affiliation{Institute for Advanced Study, 1 Einstein Drive, Princeton, NJ 08540, USA; sxu@ias.edu
\footnote{Hubble Fellow}}

\begin{abstract}
Delayed radio flares of optical tidal disruption events (TDEs)
indicate the existence of non-relativistic outflows accompanying TDEs. 
The interaction of TDE outflows with the surrounding circumnuclear medium 
creates quasi-perpendicular shocks in the presence of {toroidal} magnetic fields.
Because of the large shock obliquity and large outflow velocity, 
we find that the shock acceleration induced by TDE outflows generally leads to 
a steep particle energy spectrum, with the power-law index significantly larger than 
the ``universal" index for a parallel shock.
The measured synchrotron spectral indices of recently detected TDE radio flares are 
consistent with our theoretical expectation. 
It suggests that the particle acceleration at quasi-perpendicular shocks can be the general acceleration 
mechanism accounting for the delayed radio emission of TDEs. 
\end{abstract}


\section{Introduction}

Stars in galactic nuclei can be tidally disrupted by a central super-massive black hole (SMBH)
\citep{Ree88}.
The resulting tidal disruption events (TDEs) produce transient emission at optical/UV and X-ray bands
\citep{Sax20,Van20}.
They also eject sub-relativistic outflows, 
as indicated by the radio flares detected several months or years after the stellar disruption
\citep{Alex20,Mat21}.
The delayed radio flares are believed as the synchrotron emission from accelerated electrons when a TDE outflow interacts with the circumnuclear medium (CNM) surrounding the SMBH.
The observed synchrotron spectra provide 
constraints on the outflow properties and CNM density 
\citep{Mat21}.
As non-relativistic outflows can be ubiquitous in TDEs 
\citep{Alex16},
understanding the associated fundamental acceleration process is important not only 
for interpreting the follow-up radio observations, 
but also for determining whether TDE outflows are a possible source of high-energy cosmic rays and neutrinos
\citep{Mura20,Stei21,Wu21}.

The outflow-CNM interaction gives rise to (bow) shocks in the CNM 
\citep{Mat21}.
Shock acceleration and the energy spectrum of accelerated particles are highly sensitive to the 
obliquity angle between the upstream magnetic field and local shock normal
\citep{Jok87,NaTa95,Be21,XL22}.
{In the case of} 
dominant toroidal magnetic field in the CNM 
\citep{Silan13,Piotr17}, 
the shocks driven by TDE outflows are expected to be predominantly quasi-perpendicular with large obliquities.

Particle acceleration at quasi-perpendicular shocks is frequently considered for 
interplanetary shocks, which are mainly quasi-perpendicular at 1 AU with respect to the Parker
spiral magnetic field in the solar wind
\citep{Guo21}.
Compared with quasi-parallel shocks, quasi-perpendicular shocks have a much higher acceleration efficiency 
\citep{Jok87,XL22}.
Thus they tend to dominate the particle acceleration when there is a variation of shock obliquities 
{\citep{Deck88,FulRey90,West17,Be21}.}
\citet{XL22} (hereafter XL22) 
studied the acceleration at supernova shocks propagating through inhomogeneous media
and found that acceleration by quasi-perpendicular shocks can well explain the 
steep radio synchrotron spectra commonly seen in young supernova remnants
{(see e.g., \citealt{Mae13} for an alternative explanation)}.
The importance of particle acceleration at the quasi-perpendicular shocks driven by
TDE outflows for interpreting their synchrotron spectra  
has not been addressed.

The increasing number of radio TDEs detected in recent years 
motives this study on the origin of accelerated electrons accounting for the radio synchrotron emission. 
In this letter, we will investigate the particle acceleration process at shocks driven by TDE outflows and 
provide predictions for upcoming observations on synchrotron spectra of TDE radio flares. 
In Section 2, we describe the magnetic field geometry and shock obliquity.
In Section 3, we analyze the obliquity dependence of shock acceleration and particle energy spectrum. 
In Section 4, we compare the theoretical expectation with available measurements 
on synchrotron spectral slopes of radio TDEs. 
{Discussion and main conclusions are provided in Sections 5 and 6. }

\section{Quasi-perpendicular shocks driven by TDE outflows}
\label{sec: persho}

Possible outflows accompanying TDEs include spherical disk winds from super-Eddington accretion disks, 
unbound stellar debris,
and Newtonian jets
\citep{Mat21}.
When a sub-relativistic outflow launched by a TDE 
interacts with the CNM and the clouds around SMBHs
{\citep{Chris05,Ciur20},}
(bow) shocks form in the CNM and around the clouds
\citep{Mou22}.

As estimated from polarimetric observations
\citep{Silan13,Piotr17},
the magnetic fields in the broad line region (BLR) of active galactic nuclei (AGN) 
are dominated by the toroidal component 
due to the differential rotation in the accretion disk
\citep{Bon07}, 
with a typical strength of $\sim 10$ G.
{As SMBHs actively grow during the AGN phase, 
it can be naturally expected that 
many SMBHs still have relic toroidal magnetic fields around them
when they are in the quiescent phase. 
One example is Sagittarius A$^*$ (Sgr A$^*$), which is currently quiescent but was recently active
\citep{Nica16}.
The detailed mapping of the magnetic fields in the central parsec of the Galaxy using dust polarization
\citep{Roche18}
and the Velocity Gradient Technique 
\citep{Hu22}
reveals a significant toroidal component of
magnetic fields around Sgr A$^*$. The magnetic fields 
are associated with the circumnuclear gas that is likely spiraling into the SMBH
\citep{Roche18}.
More evidence on toroidal magnetic fields around SMBHs requires more magnetic field measurements in galaxy centers
\citep{Hu22s}. }

\begin{figure}[ht]
\centering
\subfigure[]{
   \includegraphics[width=4.1cm]{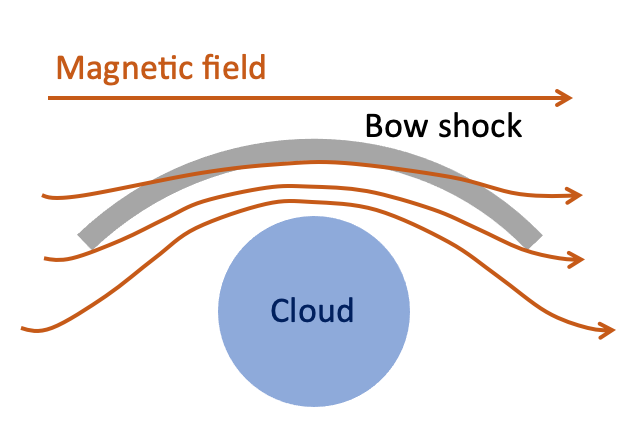}\label{fig: skebs}}
\subfigure[]{
   \includegraphics[width=4.1cm]{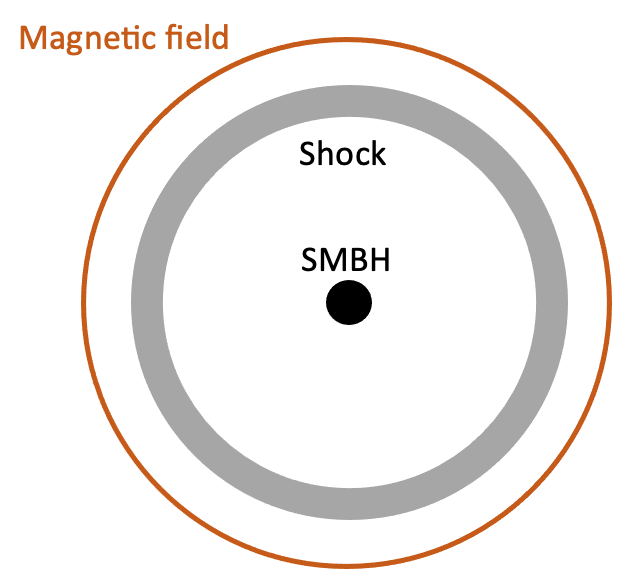}\label{fig: skess}}
\caption{Schematic illustration of quasi-perpendicular shocks driven by TDE outflows in the presence of toroidal magnetic fields
in the CNM. 
(a) A zoom-in of the bow shock around a cloud in the CNM. (b) The shock of a disk wind expanding in the CNM. 
In both cases, the shock normal is nearly perpendicular to the background mean magnetic field. 
}
\label{fig: sket}
\end{figure}

Given the toroidal geometry of the magnetic fields, 
irrespective of the specific outflow origin, 
the shocks in the CNM
are basically quasi-perpendicular, with the upstream mean
magnetic field nearly perpendicular to the shock normal. 
We illustrate the quasi-perpendicular bow shock around a cloud {near an SMBH} 
and the quasi-perpendicular shock expanding isotropically in the CNM in Fig. \ref{fig: sket}.
In the case of the bow shock, the magnetic draping leads to the development of a magnetic precursor and 
a wide opening angle of the bow shock due to the magnetic tension
\citep{Burd17}.

\section{Obliquity-dependent particle spectral index}
\label{sec: stepec}

At a subluminal oblique shock, 
due to the shock compression of magnetic fields, 
particles can be either reflected upstream 
or transmitted crossing the shock into the downstream region.
The particles undergo the 
shock drift acceleration (SDA) when they drift along the shock surface
\citep{Son69,Wu84,Arm85,BaMel01}.
In the presence of magnetic fluctuations, which can be preexisting or driven by turbulent dynamo
\citep{BJL09,Dru12,Del16,XuL17}
and accelerated particles 
\citep{Bell78,Bell2004}, 
the particles can repeatedly diffuse across the shock and are accelerated through a combination 
of SDA and diffusive shock acceleration (DSA)
\citep{Jok82,DecV86,Ostr88,Kirk89}.
The energy spectral slope of accelerated particles is determined by the 
competition between acceleration and loss of particles. 
As both processes depend on the shock obliquity, 
the resulting spectral index is also obliquity dependent.

{\it Quasi-parallel shock.}
We define 
\begin{equation}
   \beta_1 = \frac{U_1}{c \cos \alpha_1 },  ~~ \beta_2 = \frac{U_2}{c \cos\alpha_2},
\end{equation}
where subscripts $i =1$, $2$ denote the upstream and downstream regions,
$U_i$ is the shock speed in the local fluid frame
with $U_1 = 4 U_2$ for a strong shock considered in this work, 
$c$ is the light speed, 
$\alpha_i$ is the angle between the shock normal and the magnetic field, and 
$\alpha_1$ is the shock obliquity. 
At a quasi-parallel shock with a small $\alpha_1$ and a non-relativistic shock speed $U_\text{sh} = U_1$, 
we are in the limit of a small $\beta_1$.

The power-law index $\gamma$ of the particle energy spectrum 
$N(E) \propto E^{-\gamma}$ is 
\begin{equation}
  \gamma = 1- \frac{\ln P}{\ln \epsilon}, 
\end{equation}
where $E$ is the particle energy, 
$\epsilon = 1+d$, and 
$d$ is the fractional energy gain per shock crossing cycle.
$P = 1 - P_\text{esp}$ and $P_\text{esp}$ are probabilities that a particle returns to 
and escapes from the acceleration region in each shock crossing.
A shallower spectrum (i.e., a smaller $\gamma$) is expected at a larger $d$ and a smaller $P_\text{esp}$. 
For instance, 
the shallowest spectrum with $\gamma = 1$ 
corresponds to $P_\text{esp}=0$, i.e., no escape of particles.
Naturally, the spectrum becomes steep when 
$P_\text{esp}$ is large.

In the limit case of a small $\beta_1$ for a quasi-parallel shock, 
we approximately have $d\approx \beta_1$
(XL22)
and $P_\text{esp} \approx \beta_1$
\citep{Bell78},
leading to 
\begin{equation}\label{eq: orlbspec}
   \gamma \approx 1-\frac{\ln (1-\beta_1)}{\ln ( 1+ \beta_1)}.
\end{equation}
We see that $\gamma$ approaches $2$ at a sufficiently small $\beta_1$. 
We note that $\gamma=2$ is the so-called ``universal" spectral index for acceleration at a parallel shock, 
for which only the DSA is at work.

{\it Quasi-perpendicular shock.}
In a more general case with a large $\alpha_1$ and a large $\beta_1$, 
we have a general expression of $d$
(XL22)
\begin{equation}\label{eq: gendoblo}
   d = (P_{12}- P_\text{esp}) (d_{12}+ d_{21}) 
         + P_\text{esp} d_{12}
         + P_r d_r,
\end{equation}
where 
\citep{NaTa95}
\begin{equation}
    P_\text{esp} = P_{12} \frac{4\beta_2}{(1+\beta_2)^2},
\end{equation}
$P_r$ and $P_{12}$ are the probabilities of reflection and transmission,
and $d_r$ and $d_{12}$ ($d_{21}$)
are the fractional energy gain for reflection and transmission from 
upstream (downstream) to downstream (upstream).
Their expressions are provided in Appendix.

Fig.~\ref{fig: obl} shows $d$, $P_\text{esp}$ (Fig. \ref{fig: oblpd}), and $\gamma$ (Fig. \ref{fig: oblgam}) 
as a function of $\cos \alpha_1$ for sub-relativistic shocks.   
Both $d$ and $P_\text{esp}$ increase with obliquity. 
The significant energy gain at large obliquities,
with $d$ even exceeding unity, 
is mainly contributed by the reflection process in the SDA. 
The large $d$ also causes a slightly shallower spectrum compared to the quasi-parallel case. 
Very close to the largest obliquity for a subluminal shock with $\beta_1 \lesssim 1$, 
almost all particles are transmitted downstream. The lack of reflection 
causes the drop of $d$. Meanwhile, $P_\text{esp}$ approaches unity. 
As a result, the particle spectrum is drastically steepened, and we 
see in Fig. \ref{fig: oblgam}
a large $\gamma$ for a quasi-perpendicular shock.

\begin{figure*}[ht]
\centering
\subfigure[]{
   \includegraphics[width=8.7cm]{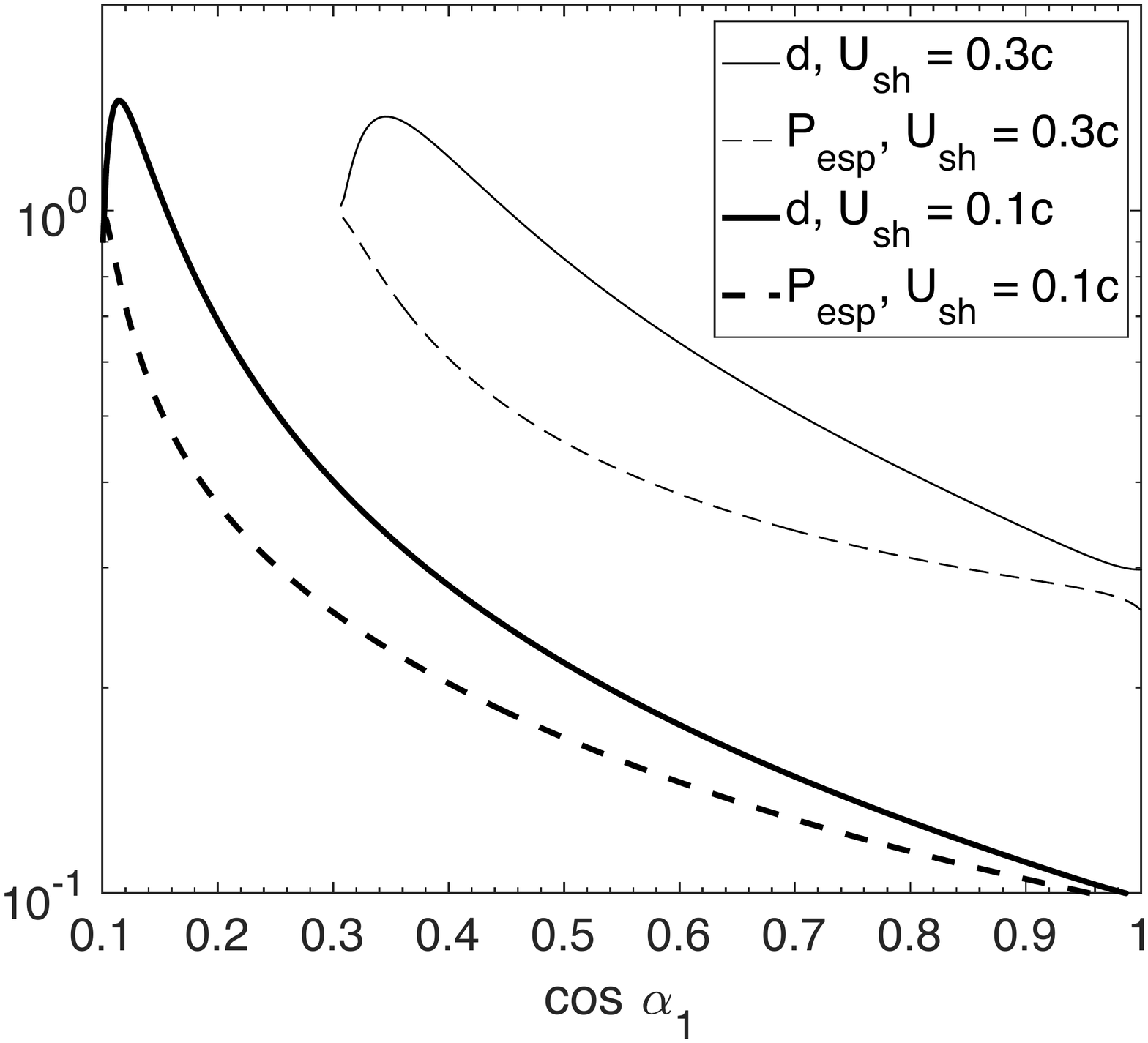}\label{fig: oblpd}}
\subfigure[]{
   \includegraphics[width=8.7cm]{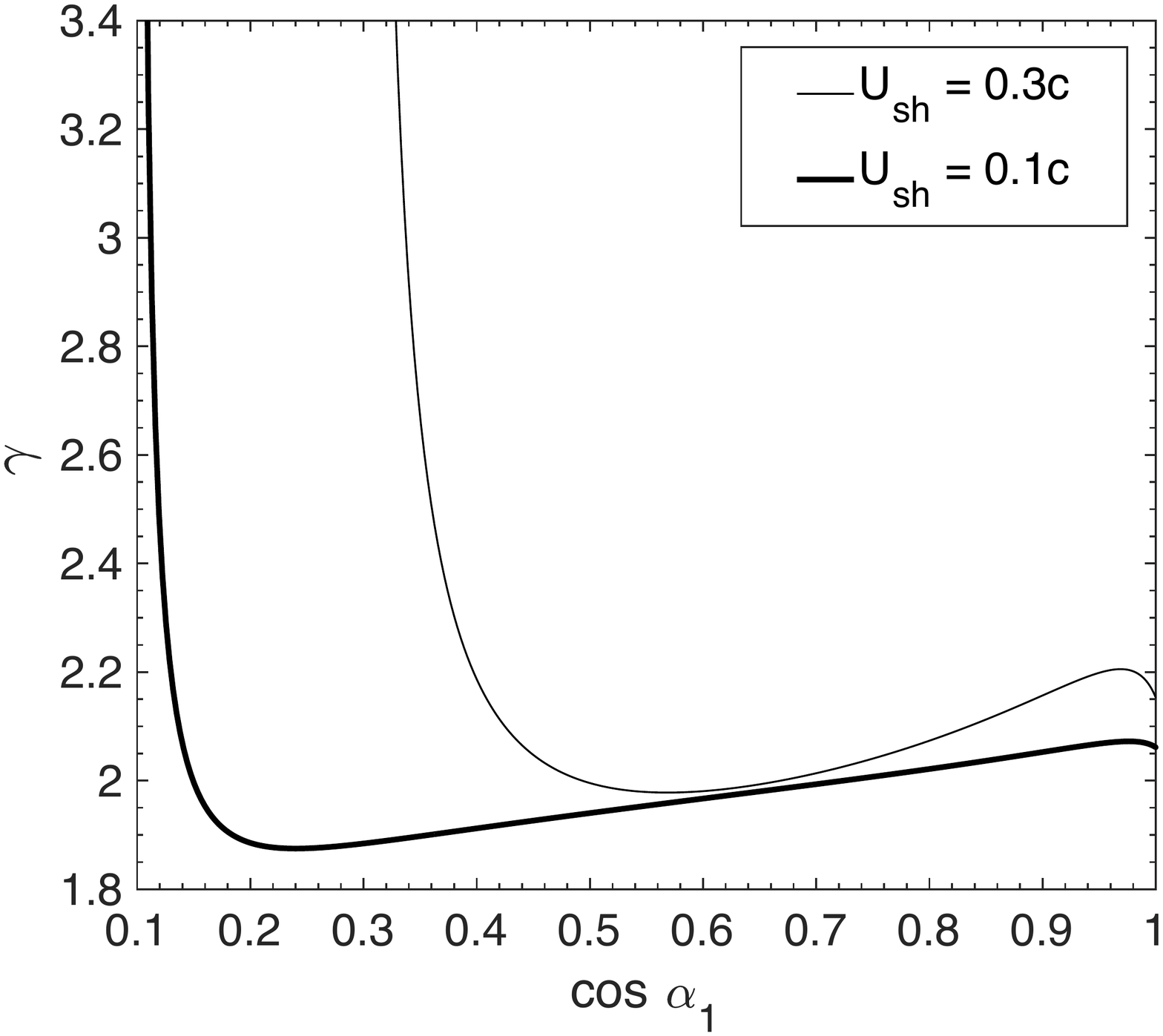}\label{fig: oblgam}}
\caption{Obliquity ($\alpha_1$) dependence of $d$, $P_\text{esp}$ in (a) and $\gamma$ in (b) for sub-relativistic shocks.  }
\label{fig: obl}
\end{figure*}

\begin{figure*}[ht]
\centering
\subfigure[]{
   \includegraphics[width=8.7cm]{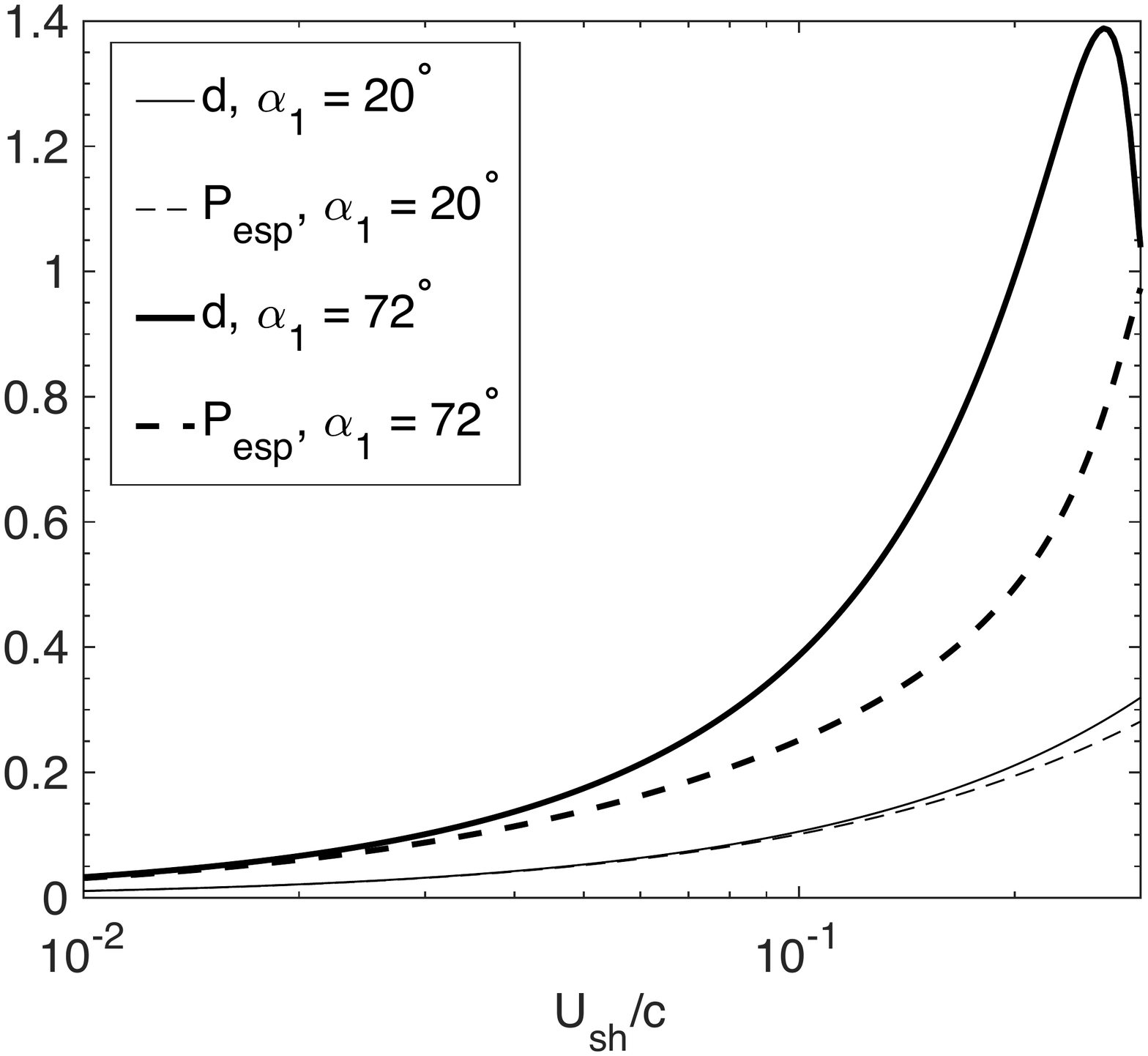}\label{fig: dpesp}}
\subfigure[]{
   \includegraphics[width=8.7cm]{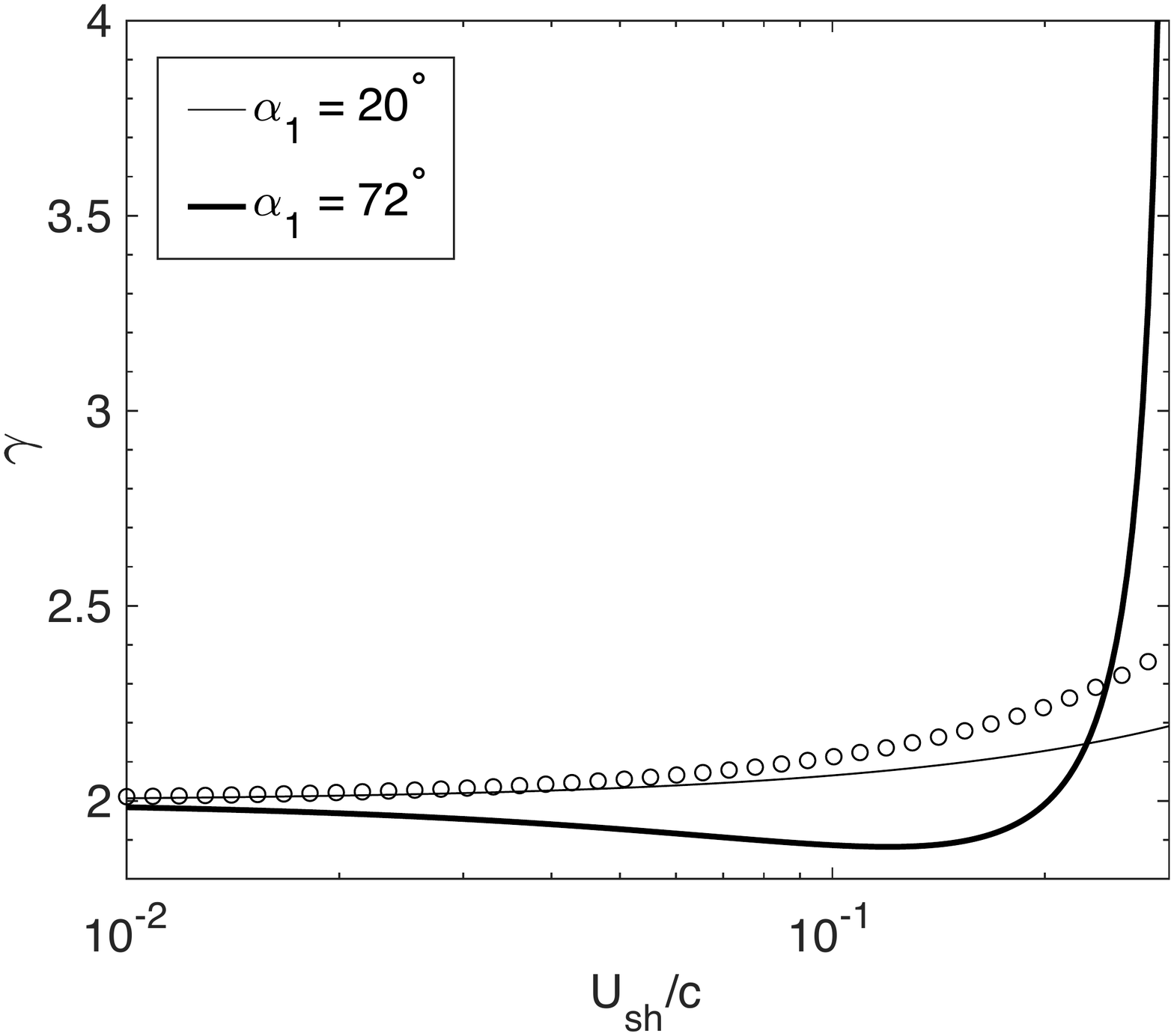}\label{fig: specind}}
\caption{$U_\text{sh}$ dependence of $d$, $P_\text{esp}$ in (a) and $\gamma$ in (b) for quasi-parallel ($\alpha_1 = 20^\circ$) and 
quasi-perpendicular ($\alpha_1 = 72^\circ$) shocks. 
Circles in (b) represent the approximate result given by Eq. \eqref{eq: orlbspec}.}
\label{fig: ush}
\end{figure*}

Fig.~\ref{fig: ush} shows $d$, $P_\text{esp}$ (Fig. \ref{fig: dpesp}), and $\gamma$ (Fig. \ref{fig: specind}) 
as a function of $U_\text{sh}$ for a quasi-parallel shock 
with $\alpha_1 = 20^\circ$ 
and a quasi-perpendicular shock with $\alpha_1 = 72^\circ$.  
We note that $\alpha_1 =72^\circ$ corresponds to $\beta_1 \approx 1$ for $U_\text{sh} = 0.3c$. 
Despite the increase of $d$ and $P_\text{esp}$ with $U_\text{sh}$, 
$d$ and $P_\text{esp}$ remain small for a quasi-parallel shock. The resulting 
$\gamma$ is well described by Eq. \eqref{eq: orlbspec} {at a small $U_\text{sh}$}, 
which is around $2$ when $\beta_1\ll 1$ and 
becomes slightly larger than $2$ at a large $U_\text{sh}$.
For a quasi-perpendicular shock, 
both $d$ and $P_\text{esp}$ have a significant increase toward a large $U_\text{sh}$. 
The drop of $d$ is still caused by the lack of reflection 
when $\beta_1$ approaches unity. 
The rapid increase of $P_\text{esp}$ at a large $U_\text{sh}$
leads to the abrupt growth of $\gamma$.

\begin{figure}[ht]
\centering
   \includegraphics[width=9.3cm]{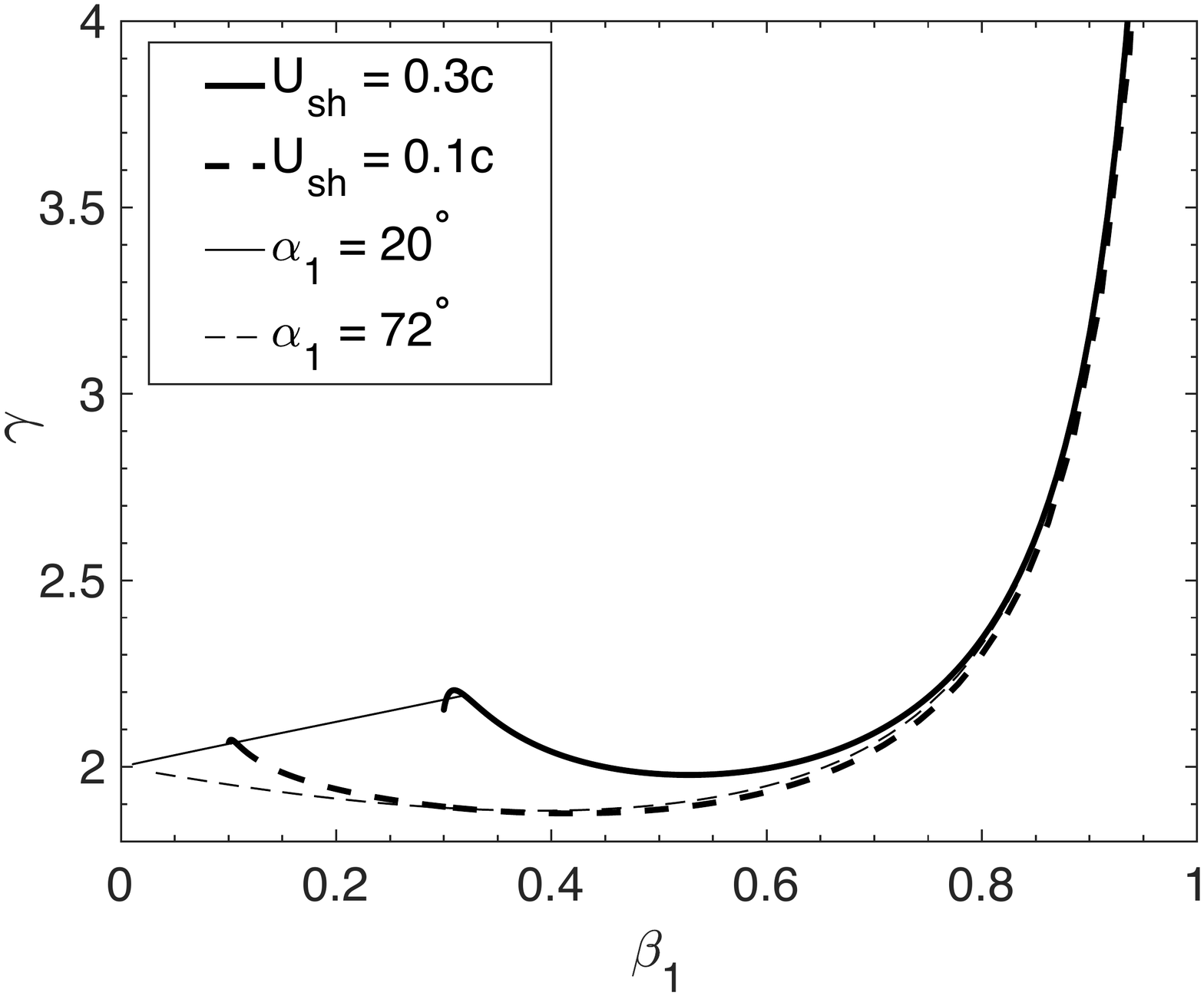}
\caption{$\beta_1$ dependence of $\gamma$ for the four cases presented in Figs. \ref{fig: obl} and \ref{fig: ush}.
Note that different symbols are used here. }
\label{fig: bet}
\end{figure}

Based on the above results, we find that a steep particle spectrum is expected for a quasi-perpendicular shock 
with both large $\alpha_1$ and large $U_\text{sh}$, that is, a large $\beta_1$. 
In Fig. \ref{fig: bet}, we present $\gamma$ as a function of $\beta_1$ for the four cases shown in Figs. \ref{fig: obl} and \ref{fig: ush}.
When a large $\beta_1$ ($\gtrsim 0.8$) can be reached, 
different cases converge toward 
a steep spectrum with a large $\gamma$.

\section{Steep synchrotron spectra of radio TDEs}

Particles are accelerated at the shocks induced by TDE outflow-CNM interaction.
Because of the combination of a large outflow velocity 
\citep{Mat21}
and a large shock obliquity (see Section \ref{sec: persho}), 
the condition of a large $\beta_1$ is satisfied.
Naturally, we expect a steep particle energy spectrum (Section \ref{sec: stepec}). 
The spectral index $s$ of synchrotron emission from accelerated electrons is related to $\gamma$ by 
$s =(\gamma-1)/2 $.
For a quasi-parallel shock in the small $\beta_1$ regime, there is $\gamma\approx 2$ and thus $s \approx 0.5$. 
A steeper particle spectrum at a large $\beta_1$ should lead to a steeper synchrotron spectrum.

\begin{figure}[ht]
\centering
   \includegraphics[width=9.3cm]{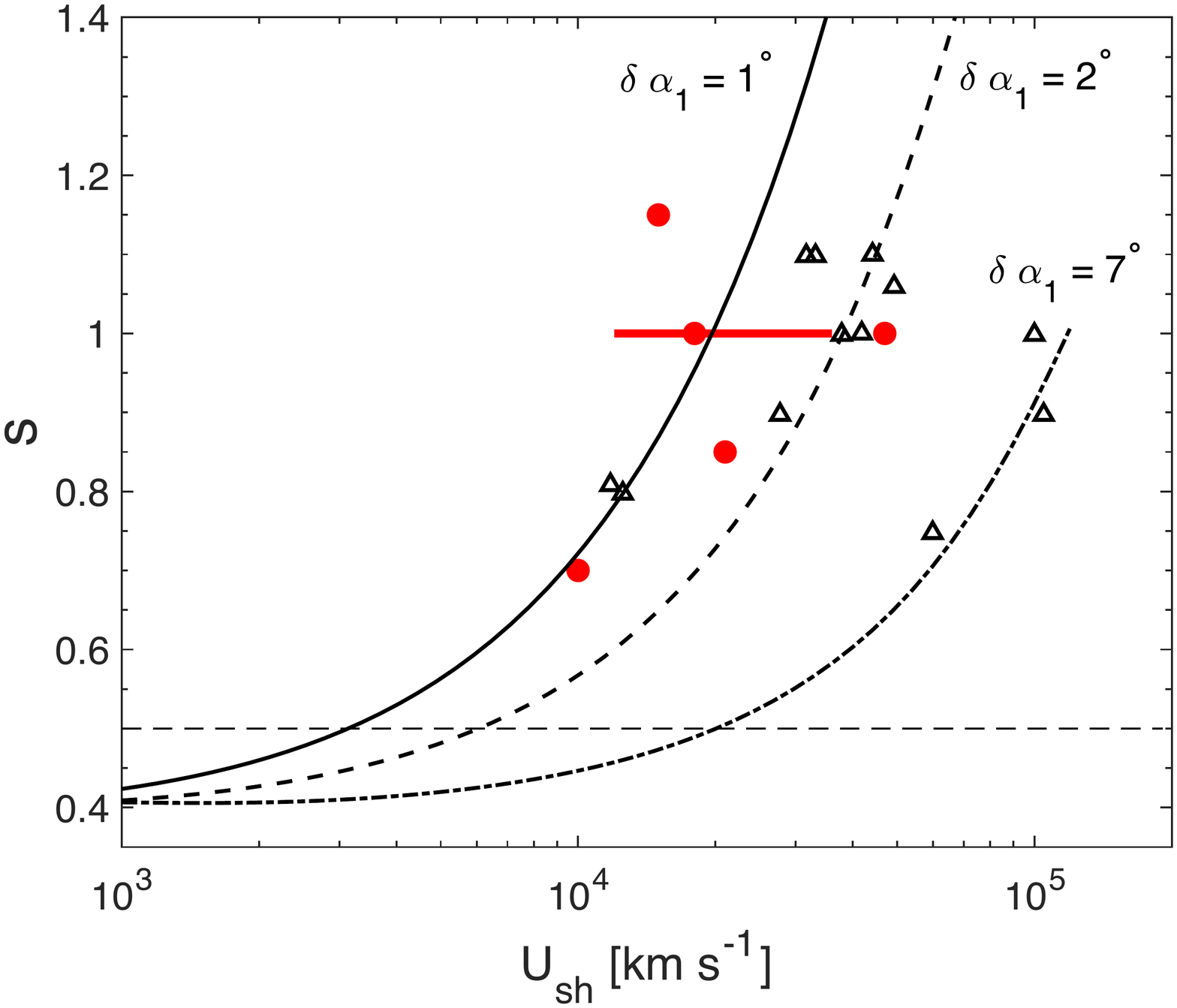}
\caption{ Radio synchrotron spectral index versus shock velocity for TDE outflows (red circles, see Table \ref{tab:tde} for the references) and 
extragalactic SNRs (triangles, see the references in \citet{Be21}).
The red line corresponds to ASASSN-14li with a range of $U_\text{sh}$. 
The three curves are taken from XL22. 
They represent the theoretical expectation for a quasi-perpendicular shock with a small obliquity variation $\delta \alpha_1$ caused by magnetic fluctuations. 
The horizontal dashed line marks $s=0.5$ for a parallel shock. 
}
\label{fig: com}
\end{figure}

We collect a number of TDEs with delayed radio flares and available measurements on $s$ from the literature 
(see Table \ref{tab:tde}), and we use their estimated outflow velocity as $U_\text{sh}$.
Note that the uncertainty range of $U_\text{sh}$ can be large, depending on the assumed outflow geometry 
(e.g., ASASSN-14li, \citealt{Alex16}),
outflow launching time, and CNM density profile
(e.g., iPTF 16fnl, \citealt{Hor21}).
As expected, 
their synchrotron spectra are steep, with $s$ significantly larger than $0.5$. 
In Fig. \ref{fig: com}, we compare the observations with the theoretical expectation for a quasi-perpendicular shock. 
The theoretical curve is taken from XL22. It corresponds to the particle spectral index of 
a quasi-perpendicular shock with a range of obliquities $[\alpha_{1,\text{max}} - \delta \alpha_1, \alpha_{1,\text{max}}]$, where $\alpha_{1,\text{max}}$ is the maximal obliquity for a subluminal shock, and 
$\delta \alpha_1$ is the obliquity variation induced by small magnetic fluctuations. We have 
\begin{equation}
   \gamma = 1 - \frac{\ln (1 - \overline{P_\text{esp}})}{\ln (1 + \overline{d})}, 
\end{equation}
where $\overline{...}$ denotes obliquity average over the range $[\alpha_{1,\text{max}} - \delta \alpha_1, \alpha_{1,\text{max}}]$.
We see that toward a large $U_\text{sh}$ in the large $\beta_1$ regime, 
$s$ rapidly rises with increasing $U_\text{sh}$. 
At a larger $\delta \alpha_1$, $s$ grows more gradually, with a weaker dependence on $U_\text{sh}$.

As a comparison, we also present the observational results of extragalactic young supernova remnants (SNRs), 
including SN 1987A, SN 1993J, and a number of Type Ib/c SNe 
(see \citealt{Be21} and references therein)
in Fig. \ref{fig: com}. 
We see a remarkable similarity between the two populations in terms of their 
fast but non-relativistic outflows and steep radio synchrotron spectra. 
{We note that an alternative explanation for steep synchrotron spectra of young SNRs was discussed in, 
e.g., \citealt{Mae13}.}
The similarity between the distribution of common non-relativistic outflows and rare relativistic jets in TDEs 
and that of common Type Ib/c SNe and rare long-duration gamma-ray bursts was earlier suggested in 
\citet{Alex16}. 
Our results indicate that steep synchrotron spectra 
can be generally seen for shock acceleration associated with non-relativistic 
outflows in transient phenomena.

\begin{table*}[!htbp]
\centering
\begin{threeparttable}
\caption{Outflow velocity and synchrotron spectral index of radio TDEs }\label{tab:tde} 
  \begin{tabular}{c|c|c|c|c|c}
   \toprule
                               &  XMMSL1 J0740-85 (1) & CNSS J0019+00 (2) & ASASSN-14li (3) & AT2019dsg (4) &  iPTF 16fnl (5)  \\
   \midrule
  $U_\text{sh}$ [km s$^{-1}$] & $10^4$  & $1.5 \times10^4$ &  $1.2\times10^4 - 3.6\times10^4$ & $2.1\times10^4$ & $ 1.8 \times 10^4$($4.7\times10^4$) \\
\hline
            s                  & 0.7 & 1.15 &  1 & 0.85 & 1  \\
   \bottomrule
\end{tabular}
     (1) \citet{Alex17}; 
     (2) \citet{And20};
     (3) \citet{Alex16};
     (4) \citet{Cende21};
     (5) \citet{Hor21}.\\
 The lower and upper limits on $U_\text{sh}$ for ASASSN-14li correspond to spherical and conical outflow geometries. 
 The two values of $U_\text{sh}$ for iPTF 16fnl correspond to different outflow launching time and CNM density profiles. 
 \end{threeparttable}
\end{table*}

\section{Discussion}

Observations suggest the existence of cloud-like objects around Sgr A*
(e.g., \citealt{Gill12}).
When a TDE outflow interacts with the inhomogeneous CNM and interstellar medium, 
the turbulence induced by density inhomogeneities 
can amplify both pre- and post-shock magnetic fields via the turbulent dynamo 
{\citep{Giac_Jok2007,BJL09,Dru12,Del16,XuL17}.
The density spectrum measured in the inhomogeneous interstellar medium is frequently much shallower than the 
Kolmogorov spectrum, and characterized by small-scale high density contrasts
(e.g., \citealt{XuZ16,XuZ17,Xup20}).
Compared to the shock propagation in a homogeneous medium with a Kolmogorov density spectrum
(e.g., \citealt{Ino13,Ji16}), 
the interaction of shocks with density fluctuations with a non-Kolmogorov 
shallow density spectrum leads to more significant corrugation of the shock surface and amplification of turbulent 
magnetic fields 
(Hu et al. in prep).
Due to both shock corrugation and turbulent dynamo, 
a range of shock obliquities is expected, depending on the spectral distribution and density contrast of upstream density fluctuations.
In the immediate vicinity of the corrugated shock surface, the magnetic field orientations aligned with the shock surface are preferentially seen, irrespective of the original shock obliquity 
(\citealt{Ino13}; Hu et al. in prep).} 
Because of the additional SDA induced by shock compression of magnetic fields 
and the suppressed diffusion of particles in the direction perpendicular to the magnetic field 
\citep{XY13}, 
a highly oblique shock has a much higher acceleration efficiency than a quasi-parallel shock 
and thus tends to dominate the particle acceleration given a complex magnetic field configuration and various shock obliquities
(\citealt{Jok87,Deck88,FulRey90}; XL22).

{TDEs have been suggested as candidates for producing high-energy protons and neutrinos
\citep{Mura20},
which is also supported by 
recently reported neutrino events associated with TDEs 
(e.g., \citealt{Stei21,Reus21,van21}).
However, with a steep spectrum of high-energy protons, it would be difficult for TDE outflows to account for the observed PeV neutrinos 
\citep{Wu21}.
It suggests that in addition to the shock obliquity, other physical ingredients, e.g., cosmic ray diffusion and feedback, 
may affect the acceleration of energetic protons. 
Alternatively, the neutrino production sites can be 
accretion disks 
(e.g., \citealt{Mur20}) 
or jets 
(e.g., \citealt{Wint21}).}

\section{Conclusions}

To understand the origin of delayed TDE radio flares detected in recently years, 
we study the particle acceleration at the (bow) shocks driven by TDE non-relativistic outflows. 
In the presence of toroidal magnetic fields in the CNM, these shocks are likely to have large obliquities.

Quasi-perpendicular shocks are intrinsically different from quasi-parallel shocks. 
Unlike the ``universal" spectral index of particles accelerated at a quasi-parallel shock,
a quasi-perpendicular shock  
has a much larger acceleration efficiency and gives rise to a steep particle energy spectrum at a large shock velocity. 
We find that 
the combination of a large shock obliquity and a large shock velocity, i.e., a large $\beta_1$, is the necessary condition 
for spectral steepening. 
This condition is naturally satisfied for shocks driven by the interaction of a TDE outflow with the CNM
and clouds around the SMBH. 
Therefore, steep particle spectra and thus steep synchrotron spectra are expected from TDE radio flares. 

Our theoretical understanding well explains the currently available measurements on synchrotron spectral slopes of radio TDEs. 
The synchrotron spectral index can be used to constrain the outflow properties and CNM density 
\citep{Mat21}.
Similar slopes of synchrotron spectra are also seen in young extragalactic SNRs with similar shock velocities, 
indicative of the same acceleration mechanism at quasi-perpendicular shocks driven by non-relativistic outflows associated with 
transient phenomena.

As non-relativistic outflows can be ubiquitous in TDEs
\citep{Alex16},
more measurements on synchrotron spectra are expected from 
the synergy between optical transient surveys and radio followups 
or blind radio surveys 
(see a review by \citealt{Alex20} for current surveys).
Current radio observations yielded detections of relatively 
low-redshift TDEs 
($z<0.05$; \citealt{Velz18}).
Radio non-detections of TDEs can be due to their low-luminosity outflows 
given the wide range of TDE radio luminosities 
\citep{Alex20}.
Next generation VLA (ngVLA) and SKA are expected to provide more synchrotron spectrum 
measurements for larger samples of TDEs at higher redshifts
\citep{Alex20}.
They will provide further tests
of our theoretical finding on the steep spectrum of particles accelerated by TDE outflows.

\clearpage

\acknowledgments
S.X. thanks Alex Lazarian, Bing Zhang, Wenbin Lu, James Stone, and Suvi Gezari for helpful discussion. 
S.X. also thanks the anonymous referee for helpful comments. 
S.X. acknowledges the support for 
this work provided by NASA through the NASA Hubble Fellowship grant \# HST-HF2-51473.001-A awarded by the Space Telescope Science Institute, which is operated by the Association of Universities for Research in Astronomy, Incorporated, under NASA contract NAS5-26555. 
\software{MATLAB \citep{MATLAB:2021}}

\appendix 

Here we provide the expressions of $d_r$, $d_{12}$, $d_{21}$, $P_r$, and $P_{12}$, 
which were originally introduced in \citet{Ostr88}. 
Their detailed derivation can be found in XL22.

The fractional energy gain in the process of reflection 
(``x" $=$ ``r"), transmission from upstream to downstream (``x" $=$ ``12"), 
and transmission from downstream to upstream (``x" $=$ ``21") is 
\begin{equation}\label{eq: dx}
    d_x = \frac{\int V_{rel} \Big(\frac{E_f}{E_0}-1\Big) d\mu}{S_x}.
\end{equation}
$E_f / E_0$ is the ratio of particle energies after and before a shock encounter. 
It has expressions 
\begin{equation}
   \Big(\frac{E_f}{E_0}\Big)_r 
   = \Gamma_1^2 (1+ 2 \beta_1 \mu +\beta_1^2)
\end{equation}
for reflection, 
\begin{equation}
\begin{aligned}
  \Big(\frac{E_f}{E_0}\Big)_{12} 
                         =  \Gamma_2 \Gamma_1  \Bigg(1+\beta_1 \mu- \beta_2  
                         \sqrt{(1+\beta_1 \mu)^2-\frac{1}{b}  \frac{1-\mu^2}{ \Gamma_1^2}} \Bigg) 
\end{aligned}
\end{equation}
for transmission from upstream to downstream, 
and 
\begin{equation}
\begin{aligned}
   \Big(\frac{E_f}{E_0}\Big)_{21} 
                            =  \Gamma_1 \Gamma_2  \Bigg(1+\beta_2\mu+ \beta_1 \sqrt{(1+\beta_2\mu)^2-b \frac{1-\mu^2}{\Gamma_2^2}}\Bigg) 
\end{aligned}
\end{equation}
for transmission from downstream to upstream, 
where $\Gamma_i$ is the  Lorentz factor corresponding to $\beta_i$, 
$\mu$ is the cosine of particle pitch angle, $b = B_1/ B_2$, 
and $B_i$ is the magnetic field strength. 
{The weight function in Eq. \eqref{eq: dx} is 
\begin{equation}
   V_{rel,1} = \frac{\mu\cos\alpha_1 + \frac{U_1}{c}}{ 1 + \frac{U_1 \mu\cos\alpha_1}{c}}, ~~
  V_{rel,2} = \frac{\mu\cos\alpha_2+\frac{U_2}{c}}{1+\frac{U_2\mu\cos\alpha_2 }{c}}
\end{equation}
in the upstream and downstream regions, respectively, }
and 
\begin{equation}
   S_x = \int V_{rel} d\mu.
\end{equation} 
The range of $\mu$ for the integral is 
\begin{equation}
   -\beta_1<\mu <\frac{\sqrt{1-b}-\beta_1}{1-\beta_1\sqrt{1-b}}
\end{equation}
for reflection, 
\begin{equation}
    \frac{\sqrt{1-b}-\beta_1}{1-\beta_1\sqrt{1-b}}    <\mu < 1
\end{equation} 
for transmission from upstream to downstream, and 
\begin{equation}
  -1< \mu < - \beta_2
\end{equation}
for transmission from downstream to upstream.

The probabilities of reflection and transmission from upstream to downstream are 
\begin{equation}
   P_r = \frac{S_r}{S_{12}+S_r}, ~~ P_{12} = \frac{S_{12}}{S_{12}+S_r}.
\end{equation}

\bibliographystyle{aasjournal}
\bibliography{xu}

\end{document}